 \documentclass[11pt]{article}
 \usepackage[left=2cm,top=1cm,right=2cm,nohead,nofoot]{geometry}
\usepackage[hang,small,bf]{caption} 
\usepackage{amsmath, amssymb, latexsym}
\usepackage{graphicx,hyperref}

\newcommand{\dotv}{ \boldsymbol{\cdot}}
\newcommand{\cross}{ \boldsymbol{\times}}

\newcommand{\curl}{ \boldsymbol{\nabla\times}}

 \newcommand{\fig}[1]{Fig.~\ref{fig:#1}}
 \newcommand{\eqn}[1]{Eq.~(\ref{eq:#1})}

 \newcommand{\A}{{\bf A}}
 \newcommand{\B}{{\bf B}}
 \renewcommand{\r}{{\bf r}}
\newcommand{\const}{\mbox{const}}

\begin{document}
\title{Relaxed plasma equilibria and entropy-related plasma self-organization principles}

\author{R.~L. Dewar$^{1,\star}$, M.~J. Hole$^{1}$, M. McGann$^{1}$,\\R. Mills$^{1}$ and S.~R. Hudson$^{2}$\\
%}
%\address{
1. Plasma and Fluids Theory Group, \\Research School of Physical Sciences \& Engineering,\\
   The Australian National University, Canberra ACT 0200, Australia \\
E-mail: robert.dewar@anu.edu.au. \\
2. Princeton Plasma Physics Laboratory, \\PO Box 451, Princeton, N.J. 08543, U.S.A.
E-mail: shudson@pppl.gov \\
$^{\star}$ Author to whom correspondence should be addressed.\\[12pt]}
%{\em Version 4-final, 3rd June 2008 Received:  / Accepted:  / Published: }}
%{\em Revised, 30 September 2008. Received:  / Accepted:  / Published: }}

\maketitle

\abstract{
The concept of plasma relaxation as a constrained energy minimization is reviewed. Recent work by the authors on generalizing this approach to partially relaxed three-dimensional plasma systems in a way consistent with chaos theory is discussed, with a view to clarifying the thermodynamic aspects of the variational approach used. Other entropy-related approaches to finding long-time steady states of turbulent or chaotic plasma systems are also briefly reviewed.
}
%\keywords{Plasma, Fusion, Relaxation, Chaos, Turbulence}

%\begin{document}
%%%%%%%%%%%%%%%%%%%%%%%%%%%%%%%%%%%%%%%%%%%%%%%%%%%%%
\section{Introduction}

\begin{figure}[htbp]
    \centering
% http://photojournal.jpl.nasa.gov/catalog/PIA04866
%	\includegraphics[height=7cm]{JupiterTokamak}
    \begin{tabular}{|c|c|}
        \hline
% http://photojournal.jpl.nasa.gov/catalog/PIA04866
 	\includegraphics[scale=0.25]{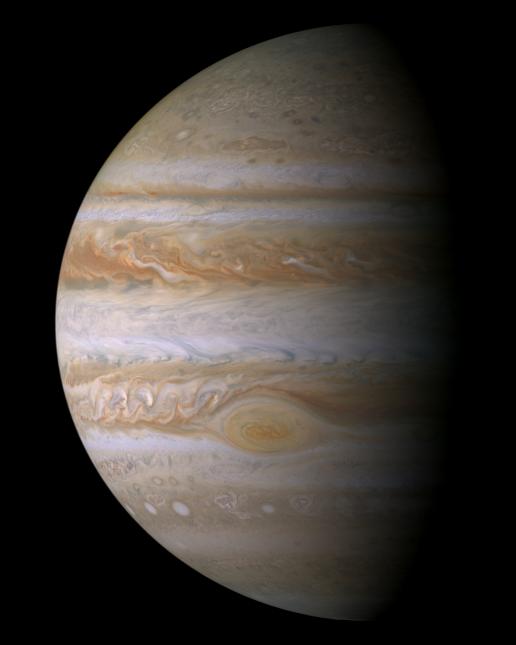}
% http://fusion.gat.com/theory/pmp/
&     
 	\includegraphics[scale=0.45]{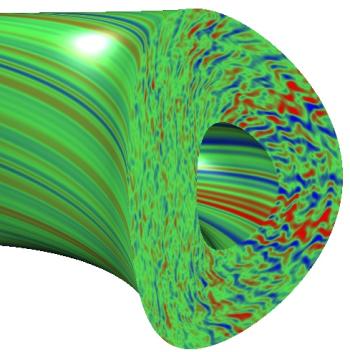}
  \\
        \hline
    \end{tabular}

    \caption{Analogue systems exhibiting self-organization in their quasi-two-dimensional turbulent dynamics:
    Left panel shows Cassini Jupiter Portrait (NASA image PIA04866),
    showing the strongly zonal nature of the solar-energy-driven turbulent
    atmospheric dynamics.  Right panel (courtesy Jeff Candy
    \url{http://fusion.gat.com/theory/Gyro}) shows a simulation of the analogous
    turbulent dynamics in a tokamak plasma, driven by heat primarily coming from the central part of the
    plasma (not shown). (N.B. the zonal direction in this case is the short way around
    the torus.}
    \label{fig:JupitervsTokamak}
\end{figure}

Magnetically confined fusion plasmas are thermodynamically nonequilibrium systems, where particles and energy are injected (or generated by fusion reactions) deep in the plasma, providing heat which flows towards the much colder edge region. This creates a kind of heat engine that drives both turbulent flows and more laminar zonal flows, somewhat analogous to the way solar energy deposition near the equator drives the dynamics of planetary atmospheres and oceans (see \fig{JupitervsTokamak}).

Confinement of strongly heated plasmas against turbulent diffusion across the magnetic field has been found, surprisingly, to improve in some circumstances due to the spontaneous formation of transport barriers related to strongly sheared zonal flows \cite{Terry_00a} driven by \cite{Diamond_Itoh_Itoh_Hahm_04,Dewar_Abdullatif_07} turbulence arising from instabilities that tap the large free energy provided by the heating and fueling of the plasma. The best-known example is the Low to High (L-H) confinement transition, where the transport barrier forms at the edge of the plasma, but internal transport barriers have been found as well. Similar sheared-zonal-flow transport barriers also occur in the atmosphere \cite{Terry_00a}, for instance at the edges of the equatorial jet and the polar vortices.

The type of plasma turbulence referred to above is driven by temperature and density gradients, causing low-frequency plasma instabilities of the \emph{drift wave} class (analogous in geophysics to Rossby waves \cite{Hasegawa_Maclennan_Kodama_79,Hasegawa_Maclennan_Kodama_79err}). These modes have little effect on the magnetic field but degrade confinement by eddy motions transporting plasma \emph{across} the magnetic field lines.

Another type of instability, of the \emph{tearing mode} class, driven by electric currents in the plasma, gives rise to electromagnetic turbulence. These modes cause magnetic reconnection, changing the topology of magnetic field lines. This effect is also potentially deleterious to plasma confinement  because transport \emph{along} magnetic field lines is very rapid. Toroidal magnetic confinement systems are designed with the intent that magnetic field lines stay on topologically toroidal surfaces (\emph{invariant tori} in the language of Hamiltonian nonlinear dynamics \cite{Arrowsmith_Place_91,Lichtenberg_Lieberman_92}), but field-line tearing can destroy such surfaces and give rise to chaotic regions that allow anomalous transport along the magnetic field lines.

Because of the complexity of these phenomena, modelling the long-time behaviour of a fusion plasma \emph{ab initio} is very difficult and various quasi-thermodynamic variational approaches \cite{Dnestrovskij_Dnestrovskij_Lysenko_05} have been proposed to predict the steady state to which a plasma will relax given some global constraints.

In Sec.~2 we review the variational principle first introduced in astrophysics by Woltjer and developed physically and mathematically in the fusion context by Taylor and other authors. We then, in Sec.~3, indicate how this approach is being extended to three-dimensional magnetic confinement systems, spelling out the (very elementary) thermodynamics involved in more detail than elsewhere in the literature. In Sec.~4 we mention very briefly other approaches that may have application in plasma physics, and point the way to future research in Sec.~5.

\section{The plasma relaxation concept}%\label{sec:Relaxation}

Although plasmas are definitely not in global thermal equilibrium, we assume that most degrees of freedom relax quickly. Thus, after an initial transient, the system reaches a statistical quasi-equilibrium characterized by only a few parameters \cite{Finn_Antonsen_83}, which evolve slowly due, \emph{inter alia}, to the smallness of flows of matter and energy between the plasma and the outside world over the short relaxation timescale. Relaxation theory describes equilibrium states of a system on an intermediate timescale, long compared with relaxation times, but short compared with heating and confinement times. Thus we take the plasma to be closed and thermally isolated and freeze the slow parameters, imposing them as constraints.

\begin{figure}[htbp]
\centering
\includegraphics[width=15cm]{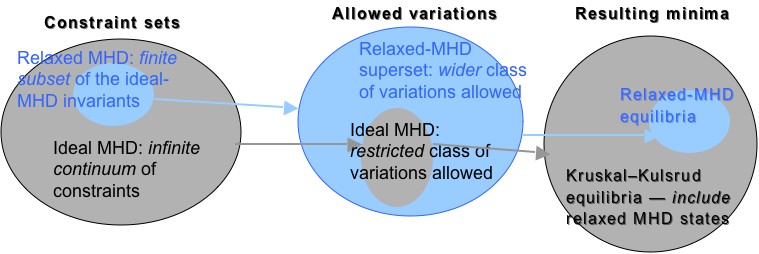}
\caption{Constraint sets, spaces of allowed variations, and equilibrium states:  Illustrating how broadening the space of allowed variation narrows the class of equilibria. }
\label{fig:Venn}
\end{figure}

We shall seek a \emph{static} equilibrium---we assume the plasma relaxes to a state with no mass flow. Also, we model the plasma as a single magnetohydrodynamic (MHD) fluid, a crude but surprisingly good approximation for the purpose of constructing background equilibrium solutions on top of which more sophisticated physics can be modelled. Finally, we use only constraints that are conserved in ideal MHD (\emph{ideal} here meaning a single-fluid, inviscid, electrically perfectly conducting, perfect gas model). The general variational theory of ideal-MHD equilibria was enunciated by Kruskal and Kulsrud \cite{Kruskal_Kulsrud_58}, basing their theory on the minimization of the total plasma energy, electromagnetic plus kinetic:
\begin{equation}
\label{eq:Energy }
	W \equiv \int_{\cal P} \left(\frac{B^2}{2\mu_0} + \frac{p}{\gamma-1}\right) d\tau \;,
\end{equation}
subject to the full set of ideal-MHD constraints. Here the plasma volume $\cal P$ is assumed to be bounded by a perfectly conducting wall, $d\tau$ denotes a volume element, $B$ is the magnetic field strength (SI units---$\mu_0$ is the permeability of free space) $p$ is the plasma pressure, and $\gamma \equiv c_p/c_v$ is the ratio of specific heats [so the internal energy of the plasma is $U = \int p d\tau/(\gamma - 1)$]\footnote{The pressure $p$ in a fusion reactor is on the order of atmospheric, while the temperature is on the order of $10^{8\,\circ}$K, so the particle density is on the order of a millionth of that in the atmosphere---insofar as local thermal equilibrium applies, the electrons and ions are ideal gases to a very good approximation.}. We likewise base our variational principle on the minimization of $W$,\footnote{Finn and Antonsen \cite{Finn_Antonsen_83} observe that, in a fully relaxed system, extremizing energy at fixed entropy is equivalent to maximizing entropy at  fixed energy. However, this is problematical in the multi-region relaxation model described in the next section as there are separately conserved entropies in each region.} and, because we use constraints that are ideal-MHD invariants, our equilibria are automatically a subset of those treated by Kruskal and Kulsrud. This approach is analogous to the Energy-Casimir method \cite[p.  511]{Morrison_98}, often called Arnold's method, and is illustrated schematically in \fig{Venn}. It is the conceptual basis for our generalization of relaxation theory discussed in the next section.

Woltjer  \cite{Woltjer_58a} observed that the ``magnetic helicity''
\begin{equation}
\label{eq:Helicity}
	K \equiv \frac{1}{2}\int_{\cal P} \A \dotv \B \, d\tau \;,
\end{equation}
where $\A$ is the vector potential such that $\B = \curl \A$, is an ideal-MHD invariant and used this as the only constraint, giving a constant-pressure equilibrium with a force-free Beltrami field,
\begin{equation}
\label{eq:Beltrami}
	\curl\B = \mu\B \;,
\end{equation}
as Euler--Lagrange equation (the constant $\mu$ being a Lagrange multiplier).

Taylor \cite{Taylor_74,Taylor_86} argued that the helicity $K$ is the ``most conserved'' invariant for a plasma in which turbulence causes field-line reconnection and showed that the Beltrami solutions modelled the results from the UK Zeta experiment well.
  
The Woltjer-Taylor relaxation approach has been generalized to two-fluid Hall MHD by Yoshida \emph{et al.} \cite{Yoshida_etal_01,Yoshida_Mahajan_02} using an additional helicity constraint involving both the magnetic field and the fluid vorticity. Also, Ito and Yoshida \cite{Ito_Yoshida_96} developed a statistical mechanical form of relaxation theory using the Shannon or R{\'e}nyi entropy, and Minardi \cite{Minardi_05} has derived the force-free relaxed state from an argument based on his magnetic entropy concept.

\begin{figure}[htbp]
\centering
\includegraphics[height=7cm]{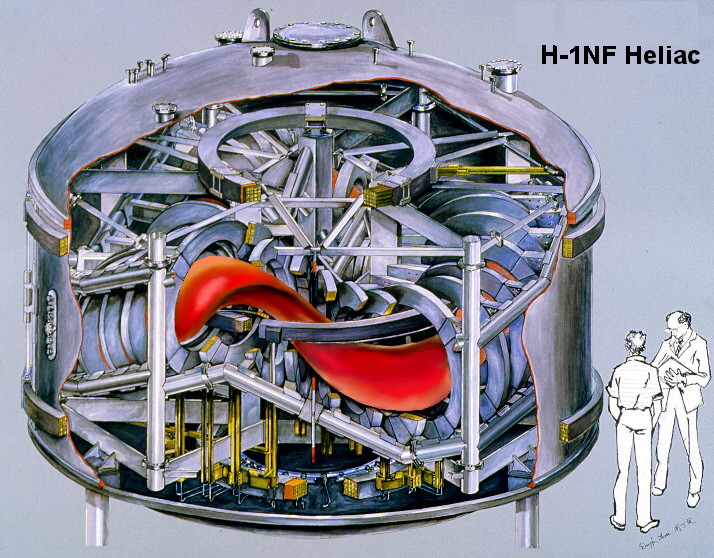}
\caption{H-1 National Facility heliac stellarator at the Australian National University. The strongly nonaxisymmetric, helically deformed toroidal plasma is indicated in red.}
\label{fig:H1}
\end{figure}

\section{Nonuniform relaxation}%\label{sec:PartialRelaxation}

The work mentioned in the previous section assumed that the plasma relaxes uniformly throughout its volume, which is both undesirable from a confinement point of view and unrealistic in all but the most turbulent experiments. To allow spatially nonuniform relaxation to be modelled, Bhattarcharjee and Dewar  \cite{Bhattacharjee_Dewar_82} expanded the set of ideal-MHD invariants used as constraints in the minimization of $W$ by taking moments of $\A\dotv\B$ with ideal-MHD-invariant weight functions that were smooth functions of position.

To be more precise, in this early work the magnetic field, which can be described as a Hamiltonian dynamical system, was assumed to be \emph{integrable}, so the plasma volume was foliated by invariant tori. Thus the weight functions were taken to be smooth functions of the flux threading these tori. However, the assumption of integrability is appropriate only for systems with a continuous symmetry (known as \emph{two-dimensional systems} because of the existence of an ignorable coordinate). This is a reasonable assumption for tokamaks, which are, neglecting the discreteness of the coils providing the toroidal magnetic field, axisymmetric. These machines rely on a large toroidal plasma current to provide the poloidal magnetic field required for confinement, and this current is a potential source of reconnection-causing instabilities, including major disruptions of the plasma.

In the class of machines known as stellarators (e.g. \fig{H1}), external coils are used instead of the toroidal plasma current, producing a more quiescent plasma but at the expense of axisymmetry. It is the development of a theory of MHD equilibria in stellarators, one which takes into account the problem of field-line chaos, that is the main motivation for our current work on finding a generalization of variational relaxation theory to three-dimensional systems.

\begin{figure}[htbp]
\centering
\includegraphics[height=7cm]{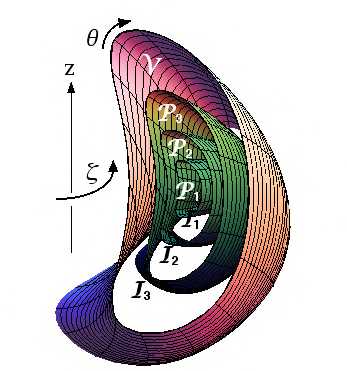}
\caption{Nested annular toroidal relaxation regions ${\cal P}_i$ and vacuum region $\cal V$ separated by KAM transport barriers, ${\cal I}_i$, as described in the text. Also shown are arbitrary poloidal and toroidal angles, $\theta$ and $\zeta$, respectively, which allow the toroidal interfaces ${\cal I}_i$ to be specified parametrically by $\r = \r_i(\theta,\zeta).$}
\label{fig:Multi}
\end{figure}

A nonaxisymmetric system is generically not integrable---there will be islands and chaotic regions in the magnetic field of a stellarator. (By \emph{chaotic} magnetic field region we simply mean a volume filled ergodically by a single field line.) Since transport along magnetic field lines, \emph{parallel transport},  is very rapid in a hot plasma \cite{Finn_Antonsen_83}, the temperature, density and pressure will rapidly become uniform in a chaotic region.

However, the Kolmogorov-Arnol'd-Moser (KAM) theorem (e.g. \cite[p.  330]{Arrowsmith_Place_91} or \cite[p. 174]{Lichtenberg_Lieberman_92}) gives reason to believe that some invariant tori survive smooth perturbation away from integrability, provided their winding number (\emph{rotational transform} in magnetic confinement jargon) is sufficiently irrational that they obey a Diophantine criterion relating to approximation by sequences of rationals\footnote{While the magnetic field can be described by a Hamiltonian, it cannot simply be written in the standard form assumed in proving the KAM theorem: $H_0+\epsilon H_1$, with $H_0$ an integrable field and $H_1$ a known perturbation. This is because plasma currents, as yet unknown, also change with geometric perturbation and they modify the Hamiltonian. Thus the KAM theorem, as normally understood, is not directly applicable to this Hamiltonian. However, generalizing previous work \cite{Berk_etal_86,Kaiser_Salat_94}, we have studied a simpler Hamiltonian problem associated with force balance across a \emph{fixed} KAM barrier that shows the rotational transforms on either side of such a surface must be strongly irrational, as in KAM theory.}. By definition, magnetic field lines cannot cross an invariant torus, so such a torus will separate chaotic regions of the plasma and be impermeable to parallel transport, allowing a pressure differential to exist between the regions. We proceed on the assumption that some invariant tori  ${\cal I}_i$ do exist  (\fig{Multi}), and, for simplicity, assume maximal chaos in the regions ${\cal P}_i$ between them, so that the pressure $p_i$ in each such region is constant. We term such pressure-jump-sustaining interfaces, which can be thought of as impermeable ideal-MHD membranes, \emph{KAM barriers}. 

Thus we have recently proposed \cite{Hudson_Hole_Dewar_07} that the generalization of the Woltjer-Taylor approach appropriate to three-dimensional geometry is the minimization of the total plasma energy
\begin{equation}
\label{eq:MultiEnergy }
	W \equiv \sum_i\left( \int_{{\cal P}_i}\frac{B^2}{2\mu_0} d\tau + \frac{p_i V_i}{\gamma-1}\right)
		+ \int_{\cal V} \frac{B^2}{2\mu_0}  d\tau
\end{equation}
subject to the helicity constraints $K_i = \const$, where $K_i$ is defined as in \eqn{Helicity} with $\cal P$ replaced by ${\cal P}_i$, and $V_i$ is the volume of region ${\cal P}_i$. The magnetic fluxes threading the ${\cal P}_i$ are conserved holonomically by restricting allowed (Eulerian) variations in $\A$ at the boundaries $\partial{\cal P}_i$ to be of the form 
\begin{equation}
    \delta \A = \delta \r_i \cross\B
             + \delta a\,{\bf n}_i + \nabla\delta\chi \quad
    \mathrm{on}\: S_i,
    \label{eq:tangentialA}
\end{equation}
where $\delta \r_i$ is  the variation in the position vector $\r = \r_i(\theta,\zeta)$ (see \fig{Multi}) of a point on the boundary, ${\bf n}_i$ is the unit normal, $\delta a$ is an arbitrary function that allows nonideal variations, and $\delta\chi$ is an arbitrary gauge term. This constraint leaves loop integrals of $\A$ as Lagrangian invariants so fluxes are conserved.

As we allow shape variations in the barrier surfaces, in addition to helicity conservation we need to constrain the pressure variations. Since we are working on the intermediate timescale, short compared with heating and confinement times, we assume the geometric variations to be both particle-number-conserving and isentropic. For an ideal (perfect) gas the entropy $S$ is given in terms of the pressure $p$ and volume $V$ by
\begin{equation}
\label{eq:Entropy}
	S = S_0 + \frac{N k}{\gamma - 1}\ln \left(\frac{pV^{\gamma}}{p_0 V_0^{\gamma}}\right) \;,
\end{equation}
where $S_0$ and $p_0V_0^{\gamma}$ are arbitrary reference values, $N$ is the number of particles and $k$ is Boltzmann's constant. Thus constancy of $N$ and $S$ implies the well-known pressure-volume relation
\begin{equation}
\label{eq:PressureConstraint}
	pV^{\gamma} = p_0V_0^{\gamma}\exp\left[(\gamma - 1)\frac{S-S_0}{Nk}\right]  = \const \;.
\end{equation}
We assume \eqn{PressureConstraint} applies to both the ion and electron gases, so the total pressure $p \equiv p_{\rm i} + p_{\rm e}$ also obeys $pV^{\gamma} = \const$, or, equivalently, $p^{1/\gamma} V= \const$\footnote{This constant is sometimes \cite{Kruskal_Kulsrud_58,Spies_03} called ``mass,'' but \eqn{PressureConstraint} shows it is a nonlinear function of both mass, $mN$, \emph{and} entropy, $S$, and therefore this terminology is best avoided.}. Thus, introducing Lagrange multipliers $\mu_i$ and $\nu_i$ for the helicity and pressure constraints respectively, our generalized relaxed-MHD equilibrium criterion is that extremizing the ``free energy''
\begin{equation}
\label{eq:FreeEnergy}
		F \equiv \sum_i\left[
		\frac{1}{2}\int_{{\cal P}_i}\left(\frac{B^2}{\mu_0} - \mu_i \A\dotv\B\right)d\tau
		+ \left(\frac{p_i}{\gamma-1} - \nu_i p_i^{1/\gamma}\right)\int_{{\cal P}_i}d\tau
		\right]
		+ \int_{\cal V} \frac{B^2}{2\mu_0}  d\tau
\end{equation}
with respect to the vector potential $\A$, the pressures $p_i$ and the barrier surfaces $\r_i(\theta,\zeta)$ gives a static equilibrium consistent with the existence of magnetic-field-line chaos between the KAM barriers. Because the constraints are a subset of the ideal-MHD constraints (\fig{Venn}), such equilibria will also be Kruskal--Kulsrud ideal-MHD equilibria.

\begin{figure}[htbp]
\centering
\begin{tabular}{cc}
\includegraphics[height=3.8cm]{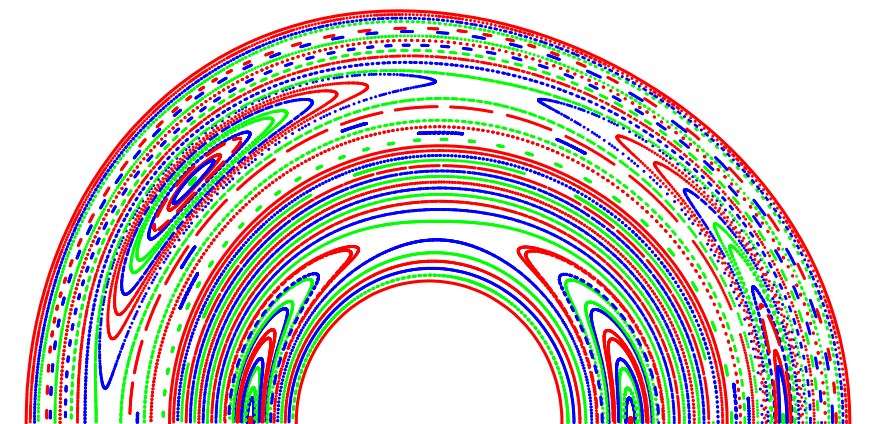} &
\includegraphics[height=3.8cm]{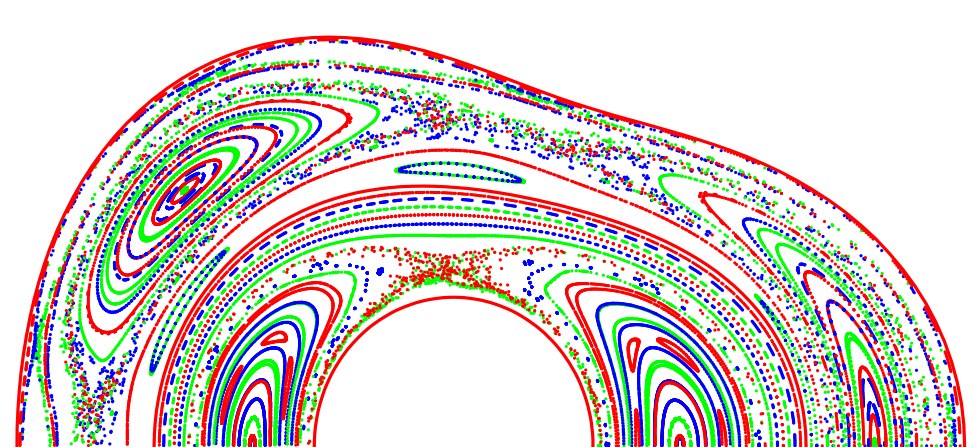}
\end{tabular}
\caption{Poincar{\'e} plots of magnetic field lines intersecting with a surface of section $\phi = \const$ for the two-relaxation-region test case described in the text, the KAM barrier being shown as a red curve topologically equivalent to the inner and outer boundaries. In the case on the left, the symmetry breaking deformation parameter $d = 0.01$, while on the right it is greater, $d = 0.04$. In the latter case the islands are clearly fatter and the chaotic regions around the island separatrices are larger.}
\label{fig:TwoRegion}
\end{figure}

The numerical implementation of this program is proceeding towards a practical 3-D equilibrium code. Figure~\ref{fig:TwoRegion} shows a two-region solution for a test case where the innermost interface is a circular-cross-section axisymmetric torus:  $R = R_0 + r \cos\theta ,\: Z = \sin\theta$,  with $R_0 = 1.0$ and $r = 0.1$, while the outermost boundary is given by $R = R_0 + r(\theta,\phi)\cos\theta.\: Z = r(\theta,\phi) \sin\theta$ with $R_0 = 1.0$ and $r = 0.3 + d \cos(2\theta - \phi) + d \cos(3\theta - \phi)$, where $d$ is an adjustable parameter which introduces nonaxisymmetry and thus chaotic fields. (In the above we are assuming a cylindrical coordinate system $R,Z,\phi$.) The (strongly irrational) rotational transforms and appropriate fluxes and pressure jump were prescribed, the Beltrami equation \eqn{Beltrami} was solved numerically in the two regions as in \cite{Hudson_Hole_Dewar_07}, and the position and shape of the KAM barrier surface was adjusted iteratively to satisfy the force balance jump condition $[\![p + B^2/2\mu_0]\!] = \const$ across it. (This relation can be derived as an Euler--Lagrange equation from \eqn{FreeEnergy}, $[\![\cdot ]\!] $ denoting the jump across the barrier surface.)

A finite element method for solving the Beltrami equation, based on the variational principle, is being developed. Also the variational nature of the problem suggests the use of gradient-based optimization methods may be better than the iterative methods so far used, but, as the constraints do not automatically keep the rotational transforms fixed at the required irrational values \cite{Hudson_Hole_Dewar_07}, care will need be taken to control the rotational transforms at the boundaries during the minimization.

The plasma will be stable not only to ideal-MHD instabilities but also to tearing and other non-ideal instabilities if the second variation of $F$ is positive definite with respect to infinitesimal perturbations respecting the constraints. The stability problem has been studied in cylindrical geometry as a generalized eigenvalue problem by defining a Lagrangian $L = \delta^2 F - \lambda N$, with $N$ a positive definite normalization. The stability condition is $\lambda \geq 0$ for all eigenvalues. Using a normalization concentrated on the ideal-MHD barrier interfaces, the perturbed field in plasma regions is computed to be Beltrami ($\curl\B = \mu\B$), with the same Lagrange multiplier $\mu$ as the equilibrium field. The interface equations between the relaxed regions produce an eigenvalue problem. 

In cylindrical geometry with axial periodicity, the displacement is Fourier decomposed, and displacements of the form $\exp i (m \theta + \kappa z)$ sought, where $m$ is the poloidal mode number, and $\kappa$ the axial wave number. Hole \emph{et al.}  \cite{Hole_Hudson_Dewar_06,Hole_Hudson_Dewar_07} have studied the stability of these  configurations as a function of mode number and number of ideal barriers, and benchmarked these results to earlier single interface studies.  Hole \emph{et al.} also revealed a singular limit problem: the relaxed-MHD stability of a two-interface plasma differs, in the limit that the two interfaces merge, from that of the corresponding single-interface plasma. 

The discrepancy has been resolved by Mills \cite{Mills_07}, who studied the stability of configurations in which the inter-barrier region was taken to be an ideal-MHD fluid rather than a relaxed region. In this case, the ideal stability of resonances in the inter-barrier region was handled explicitly, as opposed to the Woltjer--Taylor relaxed treatment,  in which resonances do not explicitly feature. Plasmas with finite-width ideal-MHD barriers showed similar stability to the single-interface configuration in the limit as the barrier width went to zero. Mills concluded it is the different treatment of resonances, which are implicit in Woltjer--Taylor-relaxed plasmas, but explicit when computing ideal-MHD stability, that is responsible for reconciling the vanishing-interface-separation paradox. In more recent work, we have also shown that the tearing mode stability threshold of the plasmas is in agreement with that found from the variational principle studied here. In ongoing work, we are also studying whether quantization in the toroidal direction leaves a stable residue of configurations in the parameter space. If so, these constrained minimum-energy states may be related to internal transport barrier configurations, which are plasma configurations with good confinement properties that form at sufficiently high heating power.

\section{Other Entropy-Related Approaches}%\label{sec:OtherApproaches}

\subsection{Maximum Entropy (MaxEnt) Principles in Plasmas and Fluids}

By the second law of thermodynamics the entropy of a closed system increases monotonically, asymptoting towards a maximum as the system approaches thermal equilibrium. Thus the application of equilibrium statistical mechanics theory can be viewed as a method for implementing the principle that systems tend towards a state of Maximum Entropy (MaxEnt).  A review by Eyink and Sreenivasan \cite{Eyink_Sreenivasan_06} traces the use of the MaxEnt principle in turbulence theory back to Onsager's work (some unpublished) in the 1940s, and cites some of the plasma and atmospheric physics literature where this approach has been used. [Onsager's equilibrium statistical mechanics is based on the Hamiltonian nature of inviscid vortex dynamics and is necessarily nondissipative. However it is very appropriate to the quasi-two-dimensional turbulence observed in strongly magnetized plasmas and planetary atmospheres (\fig{JupitervsTokamak}) where there is an inverse cascade to long wavelengths where viscous dissipation is weak.] Developments of the equilibrium statistical mechanics approach in the geophysical fluid dynamics context are further discussed in the article of Frederiksen and O'Kane \cite{Frederiksen_OKane_08} in this issue.

MaxEnt principles also occur in information theory as the least-informative estimate possible on the given information. This information-theoretic entropy concept is used in Bayesian data analysis \cite{Sivia_06} and image processing, but Jaynes \cite{Jaynes_57} also used it in physics to reinterpret statistical mechanics, with ramifications that are still being worked out today, including the Maximum Entropy Production (MEP) Principle discussed below. 

The traditional statistical mechanics approaches assume the system to be closed or in contact with a single heat bath, neither of which is appropriate to a real plasma/geophysical system where there are always fluxes of energy (and often matter) passing through the system due to heating near the centre of the plasma/planetary equator and cooling in the edge/polar regions. This problem can be overcome by the use of JaynesÕ \cite{Jaynes_57} information-theory-based generalization of statistical mechanics with intensive variable (or parameter bath) constraints; for thermodynamic systems, this is equivalent to the use of canonical-like ensembles. This, to date, has been little used in fluid mechanics (and not at all in plasma physics), though it has been used to infer steady-state velocity distributions in internal turbulent flows, such as hydraulic channels and pipes \cite{Chiu_Lin_Lu_93}.

\subsection{Maximum Entropy Production (MEP) Principles}
In recent years the idea that a nonequilibrium system develops so as to maximize its entropy production (the MEP principle) has began to attract increasing attention as a potentially powerful way to predict how a complex open system will tend to evolve. Interest in the environmental sphere was originally sparked by the work of Paltridge \cite{Paltridge_01} and the recent revival is partly due to the work of Roderick (C.) Dewar \cite{Dewar_Rod_03} using a general Jaynesian approach.

Following the recent review of entropy production principles by Martyushev and Seleznev \cite{Martyushev_Seleznev_06} we distinguish MEP in the nonequilibrium thermodynamics context from MEP in nonequilibrium statistical mechanics.

In thermodynamics the entropy of the system (or subsystem) $S$ is a state function like the internal energy U, and these must satisfy the first and second laws of thermodynamics, $\delta Q = \delta W + dU$ and $TdS \geq \delta Q$, respectively, where $T$ is the temperature, $\delta Q$ is the net heat (in whatever form) entering the system (or subsystem) and $\delta W$ the work done by the system or subsystem on the outside world (or other subsystems). If the subsystem is a small volume element and local thermal equilibrium is assumed then equality can be assumed in the second law, but globally entropy increases due to heat flows described by generalized thermodynamic fluxes and forces.

The authors of \cite{Martyushev_Seleznev_06} base their discussion of nonequilibrium thermodynamics on a principle due to Ziegler and argue that this is sufficiently general that it covers both the linear Onsager and Prigogine minimum entropy production principles, and linear and nonlinear maximum entropy production principles, reconciling them by their different interpretations.

Recently Yoshida and Mahajan \cite{Yoshida_Mahajan_08} have constructed a nonlinear thermodynamic MEP ``heat engine'' model of transport barrier formation, in a plasma or fluid system between two heat baths at different temperature, exhibiting a critical temperature difference  beyond which there is useful work $\delta W$ available to drive flows (e.g. zonal flows) that reduce turbulent transport.

The authors of \cite{Martyushev_Seleznev_06} also discuss statistical-mechanical formulations of a principle of MEP based on a principle of the most probable path in $n$-body phase space they trace back to work by Filyukov and Karpov in the late '60s. The modern approaches (e.g. \cite{Dewar_Rod_03}) also incorporate Jaynes' \cite{Jaynes_57} ideas based on Bayesian statistics. However, \cite{Martyushev_Seleznev_06} conclude that attempts to derive the MEP principle from microscopic first principles are so far unsatisfactory, as they require the introduction of additional hypotheses. They then go on to review the application of the MEP principle in different sciences.

\subsection{Minimum Entropy Production Principles}
In 1947 Prigogine proved a principle of minimum entropy production and subsequently popularized his principle and applied it in physics, chemistry and biology \cite{Martyushev_Seleznev_06}. Although its use is limited to close-to-equilibrium systems, where the thermodynamic forces and fluxes are linearly related, and it has its detractors on other grounds \cite{Barbera_99,Attard_06} a version \cite{Hameiri_Bhattacharjee_87} of minimum entropy production has had a small following in fusion plasma physics with claims of success in modelling some discharges \cite{Kucinski_98}. However, because of its linearity it is not suited to modelling the emergence of strongly nonlinear phenomena such as transport barriers.

\subsection{Minimax Entropy Production Principles}
Struchtrup and Weiss \cite{Struchtrup_Weiss_98,Struchtrup_Weiss_98r} introduce what they call a minimax principle in the context of extended thermodynamics, in which the global maximum of the local entropy production is minimized. There may also be transitional cases where entropy production is minimal with respect to some parameters and maximal with respect to others.

%%%%%%%%%%%%%%%%%%%%%%%%%%%%%%%%%%%%%%%%%%%%%%%%%%%%%%%%

\section{Conclusion}%\label{sec:Conclusion}

The multiregion-relaxation variational approach described in Sec.~3 holds strong promise of being the most satisfactory mathematical foundation on which to base the solution of the MHD toroidal equilibrium problem posed by Grad \cite{Grad_67} over forty years ago. The numerical program we are developing will not only provide a practical tool for design and analysis of fusion experiments, but will allow numerical investigation of such fundamental issues as the critical point at which a KAM barrier ceases to be able sustain a pressure jump. One hopes this will stimulate further mathematical developments beyond KAM theory (see e.g. \cite{MacKay_1992}) on the existence of KAM barriers and their breakup.

While dissipationless MHD is a standard first-cut model in fusion plasma physics because of its (relative) mathematical tractability, it is clearly inadequate to describe much of the physics of the complex self-organizing system that is a hot, toroidally confined plasma. In particular, the lack of diffusive transport  in the model allows the unphysically strong (infinite) gradients we have postulated to occur at a KAM barrier, and also more or less dictates our assumption of complete relaxation between the barriers. A first step away from this oversimplification has recently been taken by Hudson and Breslau \cite{Hudson_Breslau_08}, who used a simple anisotropic thermal diffusion model to resolve the structure of the temperature profile in a chaotic magnetic field, revealing a much more complex structure than our current relaxation model can represent. Once dissipation is present a simple energy minimization variational principle is no longer appropriate, but it may still be possible to construct a variational relaxation model of plasma steady states by using the thermodynamic MEP principle with a phenomenological Ziegler entropy production function  \cite{Martyushev_Seleznev_06}.

The Onsager MaxEnt approach has been partially explored in plasma physics, but its utility in climate modelling \cite{Frederiksen_OKane_08} suggests that it should be developed further. The use of Jaynesian MaxEnt approaches would appear to be an entirely open field in plasma physics, as is the use of statistical-mechanical MEP principles. Given the need for robust variational principles for predicting the overall behaviour of fusion we plasmas, we expect entropy-based methods to be of increasing importance in this field.

%%%%%%%%%%%%%%%%%%%%%%%%%%%%%%%%%%%%%%%%%%%%%%%%%%%%%%%%%%%%
\section*{Acknowledgements}
The first author wishes to thank the organizers of the AMSI/MASCOS Concepts of Entropy and their Applications Workshop, in particular Professor Philip Broadbridge, for the opportunity to present and discuss the topic of this paper in a stimulating environment. He acknowledges useful discussions with Drs Robert Niven and Jorgen Frederiksen. Some of the work presented was supported by U.S. Department of Energy Contract No.  DE-AC02-76CH03073 and Grant No. DE-FG02-99ER54546 and the Australian Research Council (ARC), Discovery Projects DP0452728 and DP0343765. The ARC Complex Open Systems Research Network, COSNet, grant RN0460006 also provided some support for the workshop.

\bibliographystyle{entropy}
\makeatletter
\renewcommand\@biblabel[1]{#1. }
\makeatother
\bibliography{RLDBibDeskPapers}

%%%%% Correction made on 28 February 2008 %%%%%
%\vspace{12pt}\noindent \copyright \ 2008 by MDPI (http://www.mdpi.org). Reproduction is permitted for noncommercial purposes.
%%% Original
%%% \vspace{12pt}\noindent \copyright \ 2008 by MDPI (http://www.mdpi.org). Reproduction is permitted for noncommercial purposes.

\end{document}